\documentclass[11pt]{article}

\input{FontMacros-12}

\usepackage{amsbsy}
\usepackage{amsmath}

\newenvironment{equationarraywithlabel}[1]
                          {\marginpar{\vspace{10.5mm}\hspace{-3mm}\small#1}
                          \begin{eqnarray}}{\end{eqnarray}}

\def\be#1#2\ee{\begin{equation}\label{#1}#2\end{equation}}
\def\beqa#1#2\eeqa{\begin{equationarraywithlabel}{#1}\label{#1}#2
	\end{equationarraywithlabel}}
\begin{document}
\rightline{hist-ph/0302041}
\vspace{0.5in}

\begin{center} \large {\bf P.A.M. Dirac and the Discovery of Quantum Mechanics
}\end{center}
\begin{center}\small{ Kurt Gottfried\footnote{ Newman Laboratory, Cornell
University, Ithaca NY 14853;\; kg13@cornell.edu}\\Cornell Colloquium, January 20, 2003}
\end{center}

 The occasion for this colloquium is that  Dirac was born 100
years ago last August. But why am I giving this talk, you should ask? Well, I took my
first   quantum mechanics course 51 years ago, and struggled with his great book. And
having  always been interested in the history of physics, I also browsed through Dirac's
original papers.  Ever since I have
been in deep awe of Dirac. Recently I was dismayed that no attention was being given to
his centenary, except perhaps in the UK. I was therefore both astonished and pleased to
receive an invitation from {\em Nature} last summer to write a short piece on the
occasion.\footnote{ {\em  Nature} {\bf 419}, 117 (2002)}  Being a pedant, I then spent a
lot of time reading Dirac's papers and biographical materials. {\em Nature} is aimed at a
broad scientific audience, and on one page and without equations I could not begin  to
give an understandable depiction of Dirac's accomplishments. Last fall the opportunity
arose to put what I learned into a physics colloquium. After giving the talk twice (in
Helsinki and Copenhagen) I was disatisfied, so this is a
modified attempt.

At the outset I want to make a few general remarks.
Dirac is, without doubt,  one of the great theoretical physicist in the history of
our subject; among the founders of ``modern'' physics his stature is comparable to that
of Bohr and Heisenberg, though not Einstein. Dirac had an astounding physical
intuition combined with the ability to invent  mathematics whenever he
needed it to create new physics. His greatest papers are, for long stretches,  argued
with inexorable logic, but at crucial points  there is an illogical jump into entirely
new terrain. Among the inventors of quantum mechanics,  only
deBroglie, Heisenberg, Schr\"odinger and Dirac wrote breakthrough papers that  have
such illogical and brilliantly successful long jumps.  Dirac was also a great stylist.
In my view his book belongs to the great literature of the 20th Century; it reminds me
of Kafka.   When we speak quantum mechanics we
use a language that owes a great deal to Dirac's writings.

First, a few words about Paul Dirac's life. He was born on August 8, 1902, in
Bristol. His father was Swiss and his mother from Cornwall. Dirac's father was a
domestic tyrant: he forced his children to speak to him only  in French, forbade
most social contacts, and compelled his children to pursue studies in which they
were not interested. Paul's older brother committed suicide and Paul was
exceptionally introverted and reclusive even by the standards of theoretical
physics. Dirac's difficult relationship with his father is evident from
the fact that he invited only his mother to accompany him to Stockholm when he won
the Nobel Prize in 1933.

Dirac graduated in electrical engineering from Bristol at the age of 19. He won a
scholarship to Cambridge but could not afford to accept it. He stayed on in his parent's
house, took a second degree in mathematics at Bristol, and then won a better
scholarship which allowed him to move to Cambridge in 1923 where he became a research
student of R.H. Fowler, a prominent theorist. Dirac was to stay at Cambridge until his
retirement in 1969. He then accepted a professor ship at Florida State University, and
died in Florida in 1984.

 Two recollections by his colleagues will give you some idea of his personality:
\begin{itemize}
\item Igor Tamm: He seems to talk only to children, and they have to be under ten.
\item Neville Mott: Dirac is rather like one's idea of Gandhi. He is quite
indifferent to cold, discomfort, food, etc.  ... He is quite incapable of
pretending to think anything that he did not really think. In the age of Galileo
he would have been a very contented martyr.
\end{itemize}
When Dirac won the Nobel Prize, he at first wanted to refuse it because he was afraid of
the publicity. He changed his mind after Rutherford warned him that turning it down
would produce far more publicity.

Although Dirac  focused on physics obsessively, he loved to travel and to walk in the
mountains, where he displayed exceptional endurance.  He travelled around the world
three times,  first  in 1929 together with  Heisenberg from Yellowstone to
Japan. They were both becoming quite famous by then, and the press wanted an interview
when the boat docked in Japan. Heisenberg knowing how shy his colleague was,
told the reporters Dirac was unavailable even though he was standing right beside him.

I now turn to physics. Dirac's most famous work, the Dirac equation discovered in 1928
and the prediction of anti-matter in 1931, are known to all of you. But as I have
learned, many physicists are unaware of how crucial Dirac's earlier contributions were
-- that he played a key role in the discovery and development of non-relativistic
quantum mechanics and all by himself developed  quantum electrodynamics. Given the
limited time available, I will concentrate on Dirac's early work.  I do, however, want
to give you a bird's eye view of his career, shown on this slide. I'll  comment
on most of the entries later.

\begin{itemize}
\item November 1925 -- reformulation of Heisenberg's groundbreaking paper:
canonical quantization, connection to Poisson brackets -- puts him suddenly
 at the cutting edge of theoretical physics [overlap with Born, Heisenberg and Jordan]
 \item August 1926 -- identical particles, symmetric and antisymmetric wave
functions (Fermi-Dirac statistics) [overlap with Fermi and Heisenberg]
\item December 1926 -- transformation theory -- his
favorite paper [overlap with Jordan]
\item February 1927 -- quantum theory of radiation, emission and
absorption
\item April 1927 -- scattering of light
\item January 1928 -- Dirac equation
\item December 1929 -- proposes hole theory, proton as hole
\item 1930 -- 1st edition of {\em Principles of Quantum Mechanics}
\item September 1931 -- anti-matter, in same paper as magnetic monopole
\item 1932 -- appointed Lucasian Professor of Mathematics (once held by Newton)
\item 1933 -- Lagrangian in quantum mechanics (forerunner to Feynman path integral)
\item 1933 -- Nobel Prize -- shared with Schr\"odinger
\item 1934 -- vacuum polarization, charge renormalization
\item 1937-38 -- large numbers in cosmology
\end{itemize}

For the discovery of quantum mechanics, 1925 was the watershed year. At that point the
following facts and folklore were well established:
\begin{enumerate}  \item The Bohr-Einstein relation between atomic energy levels and the
frequency of emitted radiation, $W_n-W_m = \h\om_{nm}$.
 From a classical viewpoint
this is deeply mysterious because the radiation should have the individual orbital
frequencies, not their differences.

\item Bohr's Correspondence Principle: results from quantum theory  reduce to
classical physics in limit of large quantum numbers.
\item Einstein's relations between emission and absorption rates:
\[ \Gam_{\mr{abs}} = |A_{nm}|^2\,\bar{N}_\om \qquad \Gam_{\mr{em}} =
|A_{nm}|^2\,[\bar{N}_\om +1]
\]
where $\bar{N}_\om$ is the photon spectral density and $A_{nm}$ is amplitude  of
mysterious ``virtual oscillators'' related to  the electronic motions of the initial and
final stationary states in
$n\leftrightarrow m$. Spontaneous
emission was a puzzle until Dirac's theory of
radiation  in 1927
 \item The Pauli exclusion principle and electron spin
\item  Particle-Wave duality:  advocated  by Einstein for
light since 1905 with ever stronger arguments, but controversial until discovery
of Compton scattering in 1923.  de Broglie's matter wave hypothesis of 1924 played no
role in development of quantum mechanics until Schr\"odinger's work in early 1926.
 \end{enumerate}

 Heisenberg's fundamental step in June 1925 was to dismiss the classical
concepts of {\em kinematics}, not just dynamics. Here are excerpts from the opening
of his paper:

\begin{quote}``It has become common to characterize [the] failure of the
quantum-theoretic rules as deviations from classical mechanics .. This has, however,
little meaning when one realizes that the Einstein-Bohr frequency condition already
represents such a complete departure from classical mechanics ... it seems sensible to
discard all hope of observing ... the position and period of [atomic] electrons, and to
concede that the partial agreement of the quantum rules with experience is more or less
fortuitous. [We here] try to establish a theoretical quantum mechanics, analogous to
classical mechanics, but in which only relations between observable quantities occur.
... We may pose the question in its simplest form: If instead of a classical quantity
$x(t)$ we have a quantum theoretic quantity, what quantum theoretic quantity will
appear in place of $x(t)]^2$ ?''\end{quote}

Heisenberg raised this strange question because he had long  puzzled  how to
treat the anharmonic oscillator in quantum theory, which of
course involves non-linear expressions in the kinematic variables.

In my previous attempts to tell this story, I spent considerable time on Heisenberg's
first paper, but have decided that this was unwise in a talk about Dirac because
this paper is
 very difficult to follow even in hindsight. van
der Waerden, a powerful mathematician who made significant contributions to quantum
mechanics and published an excellent  historical analysis in 1967, admits to not being
able to follow at critical points even though he had the benefit of interviews with
Heisenberg.

 So let me only say that
Heisenberg replaced any dynamical variable $A(t)$ by an array of complex
transition amplitudes
\[ A(t) \to \bra n|A|m\ket \exp[i(E_n-E_m)t/\h]\;,\]
i.e., he assumed that the time dependence was given by the Bohr-Einstein relation between
frequency and energy. (I use the term `array' because Heisenberg did not know about
matrices until Born later told him that he was using them!)  He then argued that the
array for
$AB$ is
\[ \bra n|AB|m\ket =\sum_k \bra n|A|k\ket\bra k|B|m\ket \neq \bra n|BA|m\ket\]
He showed that his scheme led to established results concerning radiative transitions
(e.g., the dipole sum rule) and the spectrum of the anharmonic oscillator.

 Heisenberg was convinced  he had done something very important, but thought
this  non-commutativity was a flaw that would have to be repaired. He was very aware of
having made logical jumps, and in private referred to this work as ``fabricating quantum
mechanics.''

Heisenberg gave a  seminar in Cambridge in July 1925, right
after he had done what I've just sketched.  He
 did not mention this work, but said something about it to
Fowler, who asked him to send proof sheets when they became available. Dirac was absent
during Heisenberg's visit. The proofs arrived at the end of August, and Fowler  gave
them to Dirac with the question ``What do you think of this? I shall be glad to
hear'' scrawled atop the first page.

 At that time the cutting edge work on the quantum theory was being done in Germany  and
Copenhagen by a small group of superbly gifted people who were in close and continuous
contact with each other: Born, Heisenberg, Kramers and Pauli  were the leading
figures, with Bohr as their father confessor.  Although Cambridge was the leading center
of experimental physics, England had contributed  little to the quantum theory. Thus
Dirac was doubly isolated -- by his own personality and by not being in the Continental
loop. When his first paper on quantum mechanics arrived in Germany it was a total
surprise; Max Born later recalled this ``as one of the great surprises of my scientific
life, for the name Dirac was completely unknown to me. ''

After pondering the Heisenberg proof sheets for two weeks, Dirac decided that its
essence was that ``that equations of classical mechanics are not in any way at fault but
that the mathematical operations by which results are deduced from them require
modification.'' Quantum mechanics, he proposed, could be inferred from classical
mechanics by redefining the canonical variables as q-numbers that obey a non-commutative
``quantum algebra,'' in contrast to ordinary numbers which he called
c-numbers.\footnote{The terms q- and c-number appear only in Dirac's next paper which I
do not mention here.} Dirac's most important new result (one that Born
\& Jordan  would not have) was that the  commutator of two q-numbers is proportional to
the q-number that corresponds to their classical Poisson bracket.  He first had this idea
while on one of his long weekly Sunday walks, but was unsure that he really remembered
what a Poisson bracket is. He then spent a restless night waiting for the library to
open. The next morning he knew  that he had  broken into  the safe.

I  will only give a very vague impression of what Dirac did in this first paper. He
started from
 \[ \sum_k\bra n|A|k\ket\bra k|B|m\ket  \]
By the correspondence principle this should have
a classical meaning for large $n$ and $m$, where  nearly diagonal terms should
dominate. By Taylor expanding about the diagonal he found
\[ (AB-BA) \to i\h \sum_i\Big(\frac{\diff A}{\diff q_i}\frac{\diff B}{\diff
p_i}-\frac{\diff B}{\diff q_i}\frac{\diff A}{\diff p_i}\Big) \]
Dirac's basic assumption was that this relation is not just valid in the
classical limit, but that  {\em   the commutator of two
q-numbers is rigorously equal to
$i\h$ times the q-number corresponding to their classical Poisson bracket.} This then
provided a complete method for constructing quantum mechanics. In particular,  if $H$ is
the Hamiltonian then any q-number
$A$ obeys the equation of motion
\[ i\h\frac{\diff }{\diff t}A(t) = A(t)H-HA(t)\;. \]
This is now called Heisenberg's equation of motion, but it does not appear in
Heisenberg first paper. The first appearance is in Dirac's paper, and in  Born \&
Jordan, though in the latter only in the representation that diagonalized $H$. Both
papers note that the Bohr-Einstein  frequency condition is a consequence.

 Dirac submitted this first paper to the Royal Society on Nov. 7, 1925, not knowing that
Born and Jordan had submitted  a paper whose most important results were
identical to his less than two weeks before.  Born \& Jordan  and Dirac both discovered
canonical quantization, with which Heisenberg's rather fragmentary scheme  was
transformed  into a self-contained and consistent  theory closely related to
classical  Hamiltonian mechanics. An amazingly detailed and extensive description of the
theory was first given in the very important paper by Born, Heisenberg and Jordan
completed in mid-November 1925.

 In January 1926,  Schr\"odinger's papers on wave mechanics started to appear.
At first Heisenberg, Dirac et al  were hostile to wave mechanics because they thought
it gave the misleading impression that the classical concepts of continuity and
 and visualizability had survived the revolution, whereas they beieved that it was a
central virtue of  their abstract theory that it did not evoke such delusions. Soon
enough -- by the summer of 1926 -- both Dirac and Heisenberg found that wave functions
were invaluable in dealing with many body problems. Their papers were the first to
recognize that indistinguishability has profound consequences in quantum mechanics that
have no counterpart whatever in classical mechanics.

Heisenberg attacked  the two-electron problem, helium, which had totally defeated the Old
Quantum Theory.  He discovered that the Pauli principle requires the two-electron wave
function to be antisymmetric, and that  the large splittings between
spin triplets and singlets was a purely electrostatic effect due to the correlations
imposed on the wave functions by antisymmetry.

Dirac, at the same time, produced a broader analysis. He showed that particles
obeying Bose-Einstein statistics must be in symmetric states while those obeying the
Pauli principle must be in antisymmetric states. Among other things, the paper has  what
later became known as a Slater determinant.   Unaware of Fermi's somewhat earlier
derivation of the Fermi distribution (which never mentioned wave functions or
antisymmetry), Dirac also derived the Fermi distribution.

On February 2, 1927, Dirac submitted to the Royal Society ``The Quantum Theory of the
Emission and Absorption of Radiation.'' This is the birth of quantum electrodynamics --
QED.  By then  about one and a half years had passes since Heisenberg's first paper.
Considering the ability of the pioneers, the speed with which they had advanced, and the
centrality of the radiation problem, this was a long wait. In any modern text on quantum
mechanics, the theory of the field and of the simplest radiative processes is among the
most straightforward chapters. So why did it take so long, and why were Dirac's
brilliant contemporaries so surprised and impressed by this paper?

The reason is that  the theoretical machine did not yet have nearly enough horsepower to
handle this problem. Those working with Schr\"odinger's equation only used
the coordinate representation, and the matrix mechanicians only used the
representation in which the Hamiltonian is diagonal. No  one had a formulation
that could handle {\em processes in which the number of degrees of freedom change.}  A
more powerful formalism not so tied to classical mechanics was needed to describe
radiation, and Dirac provide this with his  transformation theory, sent to publication
exactly two months {\em before} the radiation paper.

  The paper on transformation theory was Dirac's favorite -- he often referred to it as
``my darling,'' not the sort of word he was in the habit of using. The paper's  title is ``The physical
interpretation of the quantum dynamics'' because its central goal is the generalization
of the Born interpretation of the Schr\"odinger wave function to arbitray scalar
products between states, to use today's terms. As to the paper's form, the theory had
never before this been expressed in so elegant, general, compact and abstract a form --
the form with which we are familiar today. It is an easy read for us, but most of Dirac's
contemporaries  found it formidably abstract, and the style  did not become
popular until several decades later. Here also  $\del(x)$ is
introduced and used to great advantage:
\begin{quote} ``\ldots of course, $\del(x)$  is not a proper function of $x$, but can be
regarded only as the limit of a certain sequence of functions. All the same one can use
$\del(x)$ as though it were a proper function for practically all purposes \ldots
without getting incorrect results. One can also use [derivatives] of $\del(x)$ which are
even \ldots less ``proper'' than $\del(x)$ itself.''
\end{quote}

   Let's look at a few of the equations
from this paper:\footnote{ Dirac  did not use abstract vectors in Hilberts space, i.e.,
bras and kets,
 until much later.}
\[ (\al|A|\bet) = \int \int (\al|\xi)\: d\xi\:(\xi|A|\eta)\,d\eta
\:(\eta|\bet)\]
\[ (q'| p^n|\al) = \big(-i\h\frac{\diff }{\diff q'}\big)^n\:(q'|\al)\]
\[ \int ( q'|H|q'') \:dq''\:( q''|\al) = i\h\frac{\diff }{\diff t} \:(
q'|\al ) \]
Here $A$ is any q-number,
$\al,\bet,\xi,\eta$ the eigenvalues of $any$ set of compatible observables, whether
continuous, discrete or both, and if the system is composed of particles $q', q''$
are coordinate eigenvalues. Among other things, he thus showed that for stationary states
Schr\"odinger's wave function is the transformation function from the representation in
which the energy is diagonal to the one in which coordinates are diagonal. Furthermore,
he showed that if a system is in the state represented by the wave function
\[ \psi_\al(q') \equiv (q'|\al) \]
the probability that an arbitrary q-number $\Gam$ will display its eigenvalues $\gam$ in
some range $\gam_1\leq\gam\leq\gam_2$ is
\[ P_\psi = \int^{\gam_2}_{\gam_1} d\gam\: |(\gam|\al)|^2 \]

After hearing Dirac's first presentation of transformation theory, Heisenberg in
a letter to Pauli wrote of Dirac's ``extraordinarily grandiose
generalization'' of transformation theory.''
Jordan, I should say, also developed transformation theory at the very same time. But
transformation theory  was the last  major contribution by Dirac to be
discovered simultaneously by someone else. Starting with radiation theory, he
led the advance into relativistic quantum mechanics.

Dirac's 1927 paper on radiation theory contains: \begin{itemize} \item the first
derivation of spontaneous emission from first principles

\item  the first formulation  of second quantization
\item  the derivation of the Golden Rule of time dependent perturbation
theory\footnote{ Here Dirac devised the well-known hocus-pocus leading to a rate
proportional to time; those who have found this familiar derivation disquieting will be
pleased to learn that Heisenberg reported spending considerable effort to understanding
the result in his correspondence with Pauli.}
\end{itemize}
The paper   consists of long strings of  ``simple'' steps, interrupted
by  several logical jumps.    Again I'll
show some of the crucial equations and comment on them. He starts very innocently
with an arbitrary Schr\"odinger equation in an unspecified representation:
\[i\h \frac{\diff }{\diff t}\Psi = (H_0+V)\Psi \]
\[ (H_0-E_n)\psi_n=0\;,\qquad \Psi = \sum_n a_n(t)\,\psi_n\]
\[i\h \dot{a}_n = \sum_m V_{nm}\:a_m \qquad \mr{ (interaction ~representation)} \]
``Because'' the system is composed  of $N$ indistinguishable partices, he does not
normalize $\Psi$ to 1 but jumps to
\[ \sum_n |a_n|^2 = N\qquad |a_n|^2 = N'= 0, 1, \ldots \qquad \mbox{{\bf integers !}} \]
Then he define the new Hamiltonian
\[ \EH = \sum_{nm} a^*_nV_{nm}\,a_m\]
in which $a_n$ and $i\h a^*_n$ are canonically conjugate varibales, {\em  and still
c-numbers!} Then he  jumps to {\sc Second Quantization}: define q-numbers
\[ b_n=a_n\exp(-iE_n/\h)\qquad [b_n,b_m^{\sdag}] = \del_{nm}\]
Then the Schr\"odinger equation can be written in the representation in which the
operators $b_n^{\sdag}b_n$ are diagonal with integer eigenvalues $N'$:
\[i\h \frac{\diff }{\diff t}\:\Psi(N_1'\ldots;t) =
\sum_{nm}V_{nm}\, \sqrt{N'_n}\sqrt{N'_m+1}\:\Psi(\ldots N'_n-1\ldots N'_m+1\ldots;t)\]

The total number of particles is still conserved!  Dirac assumes
that
{ \em the  $n=0$ state where a single particle has energy and momentum zero, and is
unobservable per se, has an infinite number of photons in it, all those which
have already disappeared in absorption or that are still to appear in emission!}
This motivates the last jump:
\[ N'_0\to\infty\qquad V_{0m}\to 0 \qquad  V_{0m}\sqrt{N'_0} \to v_m
\quad(\mr{finite})\] The interaction between the radiation field and matter, he also showed, has the form
\[ V = \mr{const.}\; \sum_m \dot{X}_m \,(b_m+b_m^{\sdag})\;,\qquad \mr{i.e.,} ~~~V_{nm}=0
\quad(n,m\neq 0)\]
where $X$ is the dipole moment of the source. This is the familiar second-quantized form
of the interaction (in the dipole approximation). Use of the Golden Rule then led him to
the rates for absorption and emission, and the Einstein relations between them.

Dirac emphasized that his theory displayed the complementarity of the photon and wave
descriptions:
\begin{quote} ``There is thus complete harmony between the wave and photon description
of the interaction. We shall actually build the theory up from the photon point of view,
and show that the Hamiltonian transforms naturally into a form which resembles that for
waves.''\end{quote}
 This is an especially  important illustration of Bohr's Principle of
Complementarity, which was only formulated by Bohr later that year and reported at
the Como Conference in the fall of  1927.\footnote{Dirac wrote this paper in
Copenhagen, and it was submitted to the Royal Society by Bohr. }

It is remarkable that  while Dirac invented {\em the mathematical description} of
particle creation and destruction, he had not yet accepted {\em the concept}:
\begin{quote} ``When a light quantum is absorbed it can be considered to jump into the
zero state, and when one is emitted it can be considered to jump from the zero state to
one in which it is in physical evidence, so that it appears to have been
created.''\end{quote}
The idea that the vacuum contains an infinite number of particles
 would be used by him again in hole theory.

Two months later Dirac published the rather complicated derivation
of the Kramers-Heisenberg dispersion formula for scattering of light by extending the
theory to second order in the perturbation. And two months after that (i.e., June 1927) a
general theory of collisions, including resonance scattering, in which he derives an
expression of the Breit-Wigner type, including  a shift of the level.

 In October 1927, at the age of 25 and just two years after he first appeared on the
scene, Dirac was the youngest participant in the famous and highly exclusive Solvay
Congress where Bohr and Einstein began their long debate about the foundations of quantum
mechanics.  After a discussion at the congress with Dirac and Heisenberg of philosophy
and religion, in which Dirac expressed his distaste for such ponderings, Pauli, with his
famous acerbic whit, concluded that ``Dirac's religion is that there is no God, and
Dirac is His Prophet.''

By this time Dirac's genius was widely recognized and held in awe by his
stellar contemporaries. Not only his astonishing originality produced this attitude, but
also his  reclusive demeanor  and the extreme economy
with which he communicated.\footnote{ In a
conversation with Niels Bohr in 1958 (on BCS theory) he told me that ``Dirac is the
strangest man who ever visited my institute.'' When I asked why, Bohr explained that when
at some point he once asked Dirac what he was working on, Dirac answered that he was
trying to take the square-root of a matrix. Bohr recalled ``I wondered why such a
brilliant man is working on such a silly problem. And then, somewhat later, the proof
sheets arrived, and he had not even told me he was trying to take the square-root of the
unit matrix!!''}  Some examples of comments and recollections from that period by
physicists ranging from student to the most famous will illustrate this.
 \begin{quotation}
Christian Moller: ``... in those years we [young students]   looked into
each new issue of Proc. Roy. Soc. to see if there was a new paper by Dirac ... Often he
sat alone in the innermost room of the library in a most uncomfortable position ... He
could spend a whole day in the same position, writing an entire article, slowly and
without ever crossing anything out.''

Schr\"odinger to Bohr, during a visit to Copenhagen in the winter of 1926-27: ``Dirac has
a completely original way of thinking, which -- precisely for this reason -- will yield
the most valuable results, hidden to the rest of us. But he has no idea how {\em
difficult} his papers are for normal human beings.''

Einstein to Ehrenfest,  mid-1926: ``I have trouble with Dirac, this balancing
on the dizzying path between genius and madness is awful.''
\end{quotation}

I will  finish by  very briefly summarizing Dirac's most outstanding subsequent
contributions to physics  which, as you of course know, are more original and famous than
those that I have described:\footnote{ For a very recent discussion of Dirac's work on
General Relativity, see S. Deser, gr-qc/0301097, 1/23/003.}
\begin{itemize}

\item {\sc The Dirac Equation}, January 1928. Before then Klein, Gordon and others had
proposed a relativistic wave equation second order in time. Dirac found this
unacceptable because transformation theory had convinced him that the quantum state at
any given time suffices to specify the subsequent state. This led to him to his first
order equation. He demonstrated that it accounted for the electron's magnetic moment, and
shortly afterwards it was shown that the Dirac equation reproduces the Sommerfeld
formula for fine structure in hydrogen.

\item {\sc Hole Theory,} December 1929. All relativist equations have negative energy
solutions which cannot be removed without unacceptable consequences, as several people
proved. Dirac then proposed that his equation describes a many particle system, and that
in the ground state all negative energy levels are filled in accordance with the Pauli
principle. If a negative energy state is empty, the hole appears as a positively charged
object of positive energy, which he identified with the proton. At that time everything
was believed to be made of electrons and protons, including nuclei, and inventing new
particles was {\em  de facto verboten.}

\item {\sc Charge Conjugation.} In {\em The Principles of Quantum Mechanics} (1930), pp.
255-56, Dirac proved by the argument we use to this day that the negative energy
solutions can be transformed into positive energy solutions of {\em opposite charge and
the same mass} by a unitary symmetry transformation. He still identifies these with
protons, in the hope that interactions between the filled holes would account for the
huge mass difference. Oppenheimer showed that the lifetime of hydrogen would be
extremely short because of electron-proton annihilation; Weyl proved that the theory
must have complete symmetry between $e$ and $-e$, and concluded that the theory is
therefore in deep trouble.  Finally:

\item {\sc Anti-Matter} September 1931. {\em ``A hole, if there were one, would be a new
kind of particle, unknown to experimental physics, having the same mass and opposite
charge to an electron. We may call such a particle an anti-electron.''} \\I do not know
of any other fundamental (and successful) prediction based purely on faith in theory, for
there was not a shred of evidence that called for anti-particles. This paper  also
predicts anti-protons.

\item{\sc Magnetic Monopole.} In the same paper as anti-matter, Dirac shows that
such poles, if they exist, would compel all charges to be integer multiples of a
basic unit. This is the first proposal of a quantum states with non-trivial topolgy.

\end{itemize}

 \newpage

\begin{center}  {\bf  Bibliography}\end{center}

\noindent {\sc  Original Papers and Correspondence}
\begin{itemize}
\item {\em  The Collected Works of P.A.M. Dirac 1924-1948}, R.H. Dalitz (ed.),
Cambridge (1995)
\item W. Pauli, {\em  Scientific Correspondence,} Vol. 1, A. Hermann, K.v. Meyenn \& V.
Weisskopf (eds.), Springer (1979)
\item {\em  Sources of Quantum Mechanics}, B.L. van der Waerden, N. Holland (1967),
papers by Heisenberg, Dirac, Born \& Jordan,  (and others),
 with excellent commentary.
\end{itemize}

\noindent {\sc  Biographies}

\begin{itemize} \item R.H. Dalitz \& R.E. Peierls, {\em  Biographical Memoirs of
Fellows of the Royal Society}, {\bf 32}, 138-185 (1986)
 \item H. Kragh, {\em  DIRAC, A Scientific Biography}, Cambridge
(1990)
\end{itemize}

\noindent {\sc  Technical Histories}

\begin{itemize}
\item S.S. Schweber, {\em  QED and The Men Who Made It}, Princeton (1994)
\item A.I.
Miller, {\em  Early Quantum Electrodynamics}, Cambridge (1994)
\item S. Weinberg, {\em  Quantum Field Theory,} Vol. 1, Cambridge (1995)
\item {\em Paul Dirac -- the man and his work,} P. Goddard (ed.), Cambridge (1998)
\end{itemize}

\end{document}